\documentclass[useAMS,usenatbib,usegraphicx]{mn2e}
\usepackage[fleqn]{amsmath}

\newcommand{\der}{\mathrm{d}}
\newcommand{\R}{{R_{\rm f}}} 
\newcommand{\pk}{_{\rm pk}} 
\newcommand{\co}{_{\rm c}} 
 
\newcommand{\p}{_{\rm p}} 

\newcommand{\vir}{_{\rm vir}}

\newcommand{\hm}{_{\rm hm}}
\newcommand{\eff}{^{\rm eff}} 
 
\newcommand{\E}{_{\rm E}} 
\newcommand{\X}{_{\rm \nu}}
\newcommand{\s}{_{\rm fs}}

\newcommand{\beq}{\begin{equation}} 
\newcommand{\eeq}{\end{equation}}
\newcommand{\beqa}{\begin{eqnarray}}
\newcommand{\eeqa}{\end{eqnarray}}

\newcommand{\ti}{t_{\rm i}}
\newcommand{\DM}{_{\rm DM}}
\newcommand{\M}{_{\rm M}}
\newcommand{\CDM}{_{\rm CDM}}
\newcommand{\WDM}{_{\rm WDM}}
\newcommand{\modot}{M$_\odot$\ } 
\newcommand{\modotc}{M$_\odot$}
\newcommand{\apj}{ApJ}
\newcommand{\mnras}{MNRAS}
\newcommand{\apjl}{ApJ Lett.}

\newcommand{\jcap}{JCAP}
\newcommand{\prd}{Phys. Rev. D}
\newcommand{\lav}{\langle}
\newcommand{\rav}{\rangle}
\newcommand{\srho}{\lav\rho\rav}

\begin{document}

\title[WDM halo density profile]{Typical density profile for warm dark matter
  haloes}

\author[Vi\~nas et al.]{Jordi Vi\~nas\thanks{E-mail: jvinas@am.ub.es}, Eduard Salvador-Sol\'e and Alberto Manrique\\
Institut de Ci\`encies del Cosmos, Universitat de
Barcelona (UB--IEEC), Mart{\'\i} i Franqu\`es 1, E-08028 Barcelona, Spain}

%% Abstract and keywords

\maketitle
\begin{abstract}
Using the model for (bottom-up) hierarchical halo growth recently
developed by \citet{Sea12}, we derive the typical
spherically averaged density profile for haloes with several relevant
masses in the concordant warm dark matter ($\Lambda$WDM) cosmology
with non-thermal sterile neutrinos of two different masses. The
predicted density profiles become flat at small radii, as expected
from the effects of the spectrum cutoff. The core cannot be
resolved, however, because the non-null particle velocity yields the
fragmentation of minimum mass protohaloes in small nodes, which
invalidates the model at the corresponding radii.
\end{abstract}



\section{Introduction}

Matter in the Universe is predominantly dark and clustered in haloes
that grow through mergers and accretion. The concordant $\Lambda$,
cold dark matter (CDM), model recovers the observed large-scale
properties of the Universe: it correctly predicts the microwave
background radiation anisotropies \citep{Km11} and galaxy clustering
\citep{Col05}. However, some problems arise in the small-scale regime:
it predicts excessive substructure with a deficient distribution of
maximum circular velocities \citep{Klea99,Mooea99,Boy11} and a sharp
central cusp in the halo density profile, apparently in conflict with
the profiles observed in dwarf galaxies \citep{Goe06}. While the
disagreement in the satellite abundance and characteristics might be
explained through the inhibition of star formation owing to several
baryonic feedback processes, the cusp problem might be insurmountable
(but see \citealt{Macea12a,Govea12}).

Several modifications of the $\Lambda$CDM model have been proposed
that, keeping its right predictions at large scales, may improve those
compromised at small scales. This includes the cosmological models
dominated by self-interacting dark matter \citep{SS00,Kapea00,Bea00}
and dissipationless collisionless warm dark matter (WDM). The best
candidate particles in the latter category of dark matter are the
gravitino \citep{Eea84,HD00,Kapea05,Gea08} and non-thermal sterile
neutrino \citep{DW94,HD00,SK06}. Specifically, the case of light
sterile neutrinos is being attracting growing interest as this kind of
particles is naturally foreseen in a minimal extension of the Standard
Model.

The non-negligible WDM particle velocities at decoupling introduce a
cutoff in the power-spectrum due to free-streaming and, at the same
time, a bound in the fine-grained phase-space density. The former
effect should inhibit the formation of haloes below the corresponding
free-streaming mass scale, $M\s$. Thus, if the cusp in CDM haloes
arises from very low-mass, extremely concentrated, halo ancestors, the
absence of haloes with masses below $M\s$ would translate into the
formation of a core. The latter effect should give rise to an upper
bound in the coarse-grained phase-space density of haloes resulting
from virialisation, which could also lead to a non-divergent central
density profile. Both aspects depend, however, critically on the
poorly known way haloes fix their density profiles.

In fact, $N$-body simulations do not seem to confirm such
expectations: WDM haloes show similar density profiles as CDM haloes
(e.g. \citealt{CDW96,Cea00,Kn02,WW09,SSMM11}). Only a small hint
towards increased scaled radii has been found \citep{BOT01}), although
the opposite trend, namely a slight inflection towards sharper density
profile at small radii ($\sim 0.02 - 0.03$ times the virial radius
$R\vir$) has also been reported \citep{Cea08}. A core has only
  been found in the recent work by \citet{Macea12b}. The situation
is further complicated by the fact that simulations of WDM cosmologies
find a substantial amount of haloes with masses considerably smaller
than $M\s$. These low-mass haloes are spurious due to the periodical
grid used in simulations \citep{WW07}, but even so they could
affect the density profile of more massive haloes formed from their
mergers and accretion. In addition, the expected size of the core (if
any) for the relevant WDM particle masses is close to the resolution
of current simulations, which might explain the negative results
above.

All these uncertainties would disappear if the halo density profile
could be inferred analytically down to arbitrarily small radii
directly from the power-spectrum of density
perturbations. \citeauthor{Sea12} (\citeyear{Sea12}, hereafter SVMS)
have recently built a model for the inner structure of dissipationless
collisionless dark matter haloes in (bottom-up) hierarchical
cosmologies\footnote{In WDM cosmologies there is a minimum halo mass,
  but haloes still form hierarchically through the merger of less
  massive objects previously formed and their accretion of diffuse
  matter.} that fills this gap.

In the present Letter, we apply the SVMS model to the $\Lambda$WDM
cosmology with 2 keV thermalised and non-thermalised sterile
neutrinos. The linear power-spectrum in WDM cosmologies endowed with
non-thermal sterile neutrinos with mass $m\X$ is given by
\citep{Vea05}
\beq
P\WDM(k)=T\s^2(k)P\CDM(k)\,,
\label{transfer}
\eeq
where $P\CDM(k)$ is the power-spectrum for the $\Lambda$CDM cosmology,
given here by the wmap7 concordance model \citep{Km11}, and $T\s(k)$
is the `transfer' function, well-fitted by
\beq
T\s(k)=[1+(\alpha k)^{2\mu}]^{-5/\mu}\,,
\eeq
with $\mu=1.12$ and
\beq 
\alpha=0.1655\left(\!\frac{h}{0.7}\!\right)^{\!0.22}\!\!\left(\frac{m\X}{1~
  {\rm{keV}}}\right)^{\!-0.136}\!\left(\!\frac{\Omega\WDM}{\!0.228}\!\right)^{0.692}
\,{\rm Mpc}\,,
\label{alpha}
\eeq
being $h$ the current value of the Hubble parameter in units of 100
Mpc$^{-1}$ km s$^{-1}$, and $\Omega\WDM$. The spectrum for fully
thermalized particles with mass $m\WDM$ is essentially equal to that
for non-thermalized particles with mass $m\X$ given by \citep{Vea05}
\beq 
m\X=4.286 \left(\frac{m\WDM}{1~{\rm{keV}}}\right)^{4/3}
\left(\!\frac{h}{0.7}\!\right)^{\!-2/3}\!\!\left(\!\frac{\Omega\WDM}{0.273}\!\right)^{\!-1/3}
\,{\rm keV}\,.
\eeq
Therefore, the case of thermalised sterile neutrinos with $m\WDM=2$
keV is equivalent to that of non-thermalised ones with $m\X=10.8$
keV. The two masses, $m\X=10.8$ keV and $m\X=2$ keV here considered
yield (eq.~[\ref{alpha}]) $\alpha=0.032$ Mpc and $0.151$ Mpc,
respectively. We remind that $m\X=10.8$ keV (or $m\WDM=2$) sterile
neutrinos are compatible with observational constraints such as the
Lyman-$\alpha$ forest and the abundance of Milky Way satellites, while
the case of lower particle masses is unclear
(e.g. \citealt{Bea09,PR11}).

\section{The Model}\label{method}

In any {\it bottom-up hierarchical cosmology}, haloes form through
either major mergers or smooth accretion (including minor
mergers). Both processes involve the virialisation of the halo each
time the mass increases. As virialisation is a relaxation process
yielding the memory loss of the past history, the density profile for
haloes having suffered major mergers is indistinguishable from that
for haloes having grown by pure accretion (PA; see SVMS for a complete
rigorous explanation). Consequently, we have the right to concentrate
in this latter kind of haloes.

As shown in SVMS, {\it accreting dissipationless collisionless dark
  matter haloes} evolve from the inside out, keeping their
instantaneous inner structure unaltered. In these conditions, the
radius of the sphere with mass $M$ is exactly given by
\beq 
r(M)=\frac{3GM^2}{10\,|E\p(M)|}\,,
\label{mr2}
\eeq
where $E\p(M)$ is the total energy of the sphere encompassing the same
mass $M$ in the (spherically averaged) seed, namely a peak in the
primordial random Gaussian density field filtered at the scale
$M$. Thus, provided the energy distribution in peaks is known,
equation (\ref{mr2}) is an implicit equation for the halo mass profile
$M(r)$.

We must remark that equation (\ref{mr2}) is only valid provided the
isodensity contours in the seed reach turnaround without
shell-crossing at increasingly larger radii (see SVMS). This is
certainly the case when the initial peculiar velocities are
negligible, as those induced by random Gaussian density fluctuations,
and the seed expands in linear regime. However, if there are peculiar
velocities of non-gravitational origin such that they dominate the
dynamics at small enough scales, then the system may not expand in
linear regime and there may be shell-crossing before turnaround, so
equation (\ref{mr2}) may no more be valid. We will come back to this
possibility in Section \ref{warm}.

In the parametric form, $E\p(M)$ is given by the total energy in the
sphere with radius $R\p$ centred at the peak,
\beqa 
E\p(R\p)=4\pi\int_0^{R\p}\der r\p\, r\p^2\,\lav \rho\p\rav(r\p)~~~~~~~~~~~~~~~~~~~~~~~~~\nonumber\\
\times\left\{\frac{\left[H(\ti) r\p-v\p(r\p)\right]^2}{2}+\frac{\sigma^2\DM(\ti)}{2}-\frac{GM(r\p)}{r\p} \right\},
\label{E1}
\eeqa
together with the mass of the sphere,
\beq 
M=4\pi\int_0^{R\p} \der r\p\, r\p^2\, \lav \rho\p\rav(r\p)\,,
\label{M1}
\eeq 
where $\lav\rho\p\rav(r\p)$ is the spherically averaged (unconvolved)
protohalo density profile, $M(r\p)$ the corresponding mass profile,
$H(\ti)$ the Hubble parameter at the cosmic time $\ti$ when the seed
is considered, $\sigma\DM(\ti)$ the adiabatically evolved particle
velocity dispersion of non-gravitational origin\footnote{The velocity
  dispersion due to random density fluctuations is several orders of
  magnitude smaller and can be safely neglected (see SVMS).}  and
$v(r\p)$ is, to leading order in the deviations from spherical symmetry, the
peculiar velocity at the radius $r\p$ in the seed owing to the inner
mass excess\footnote{In equation (\ref{vp}) we have taken into account
  that the cosmic virial factor $f(\Omega)\approx \Omega^{0.1}$ is at
  $\ti$ very approximately equal to one.},
\beq
v\p(r\p)=-\frac{2G\left[M(r\p)-4\pi r\p^3\bar\rho(\ti)/3\right]}{3H(\ti)r\p^2}\,,
\label{vp}
\eeq
being $\bar\rho(\ti)$ the mean cosmic density at $\ti$. 

The steps to be followed are thus the following ones: 1) determination
of the spherically averaged protohalo density profile
$\lav\rho\p\rav(r\p)$ from the linear power-spectrum of the cosmology
considered (see Sec.~\ref{seed}), 2) calculation from it of the energy
distribution $E\p(M)$ (eqs.~[\ref{E1}]--[\ref{M1}]), and 3) derivation
of the typical halo mass profile $M(r)$ by inversion of equation
(\ref{mr2}), and of the typical spherically averaged halo density
profile, through the trivial relation
\beq
\srho(r)= \frac{1}{4\pi r^2}\frac{\der M}{\der r}\,.
\label{rhot}
\eeq

\section{Protohalo Density Profile}\label{seed}

In PA, every halo ancestor along the continuous series ending at the
halo with $M$ at $t$ also arises from one peak in the random density
field at $\ti$, filtered at the mass scale of the ancestor. Thus, the
value at the centre of the protohalo ($r\p=0$) of the convolution of
the spherically averaged density contrast profile for the protohalo,
$\lav\delta\p\rav(r\p)$, by a Gaussian window of every radius $\R$
must be equal to the density contrast of the peak (in the density
filed equally convolved),
\beq
\delta\pk(\R)=\frac{4\pi}{(2\pi)^{3/2}\R^3}
\int_0^{\infty}\der r_p\, r_p^2\,\delta_p(r_p)\,{\rm e}^{-\frac{1}{2}\left(\frac{r_p}{\R}\right)^2}\,.
\label{dp1}
\eeq 
Therefore, provided the peak trajectory, $\delta\pk(\R)$, associated
with the accreting halo were known, equation (\ref{dp1}) could be seen
as a Fredholm integral equation of first kind for
$\lav\delta\p\rav(r\p)$. Such an equation can be solved (see SVMS for
details), so this would lead to the density profile
$\lav\rho\p\rav(r\p)$ for the seed of the halo evolving by PA.

Furthermore, in any {\it random Gaussian density field}, the typical
peak trajectory $\delta\pk(\R)$ leading to a purely accreting halo
with typical density profile is the solution of the differential
equation (see SVMS and references therein)
\beq
\frac{\der \delta_{pk}}{\der \R}= -x_e(\delta_{pk},\R)\sigma_2(\R)\R\,,
\label{dp2}
\eeq
where $\sigma_2(\R)$ is the second order spectral moment and
$x_e(\R,\delta\pk)$ is the inverse of the average inverse curvature
$x$ (minus the Laplacian over $\sigma_2$) of peaks with density
contrast $\delta\pk$ at the scale $\R$ (see SVMS for the explicit form
of these two functions). The quantities $\sigma_2(\R)$ and
$x_e(\R,\delta\pk)$ depend on the power-spectrum of the particular CDM
or WDM cosmology considered, so does also the typical peak trajectory
solution of equation (\ref{dp2}). This equation can be solved for the
boundary condition $\delta\pk[\R(M)]=\delta(t)$ leading to the halo
with $M$ at $t$ according to the one-to-one correspondence between
peaks and haloes given by \citet{MSS95}
\beq
\R(M)=\frac{1}{q}\left[\frac{3M}{4\pi\bar\rho(\ti)}\right]^{1/3}~~~~~~~~ 
\delta(t)=\delta\co(t) \frac{G(\ti)}{G(t)}\,,
\label{deltat}
\eeq
where $q$ is the radius, in units of $\R$, of the collapsing cloud
with volume equal to $M/\bar\rho(\ti)$ associated with the peak,
$G(t)$ is the cosmic growth factor and $\delta\co(t)$ is the critical
linearly extrapolated density contrast for spherical collapse at
$t$. In the underlying $\Lambda$CDM cosmology here considered, such a
correspondence is given by $q=2.75$ and $\delta\co(z)=1.82+(6.03-0.472
z+0.0545 z^2)/(1+0.000552 z^3)$ (SVMS).

\begin{figure}
 \includegraphics[scale=0.44]{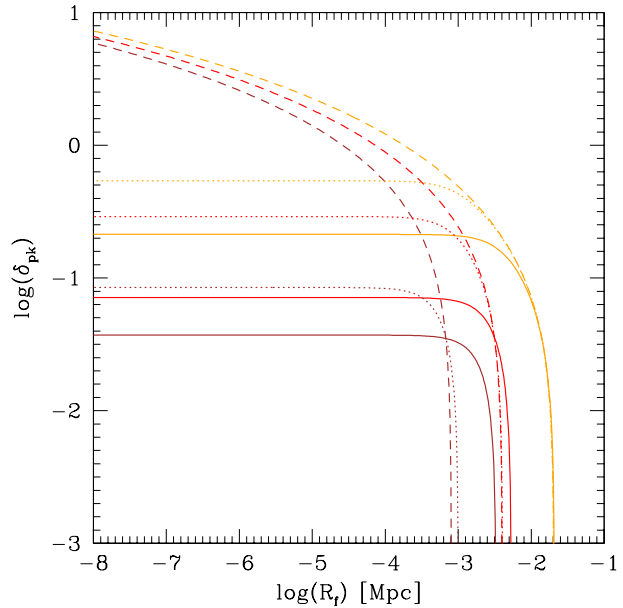}
\vskip -0.5 cm
 \caption{Central density contrast of peaks in the filtered density
   field at $z=100$ giving rise by PA to current haloes with masses
   equal to $10^9$ \modot (brown lines), $10^{11}$ \modot (red lines)
   and $10^{13}$ \modot (orange lines) as a function of the Gaussian
   filtering radius $\R$. The different curves correspond to the
   $\Lambda$CDM concordance cosmology (dashed lines) and the
   $\Lambda$WDM cosmology with $m\X=2$ keV (solid lines) and
   $m\X=10.8$ keV (dotted lines) sterile neutrinos. Filtering radii,
   $\R$, are in physical units.}
\label{f1}
\end{figure}

The peak trajectories, $\delta\pk(\R)$, solution of equation
(\ref{dp2}) in the WDM (CDM) cosmology for haloes with several masses
are shown in Figure \ref{f1}. As can be seen, the $\delta\pk(\R)$
trajectories in the WDM cosmology level off, contrarily to those in
the CDM cosmology, at some value $\delta\s$ that depends on the halo
mass. The time $t\s$ corresponding to $\delta\s$ (eq.~[\ref{deltat}])
marks the time when the first ancestor of the halo forms and initiates
the continuous series of ancestors leading by PA to that final
halo. Before that time there is no ancestor of halos with that mass in
the WDM cosmology. Instead, there are halo ancestors down to any
arbitrarily small time in the CDM cosmology, with no spectrum cut-off.

\begin{figure}
 \includegraphics[scale=0.44]{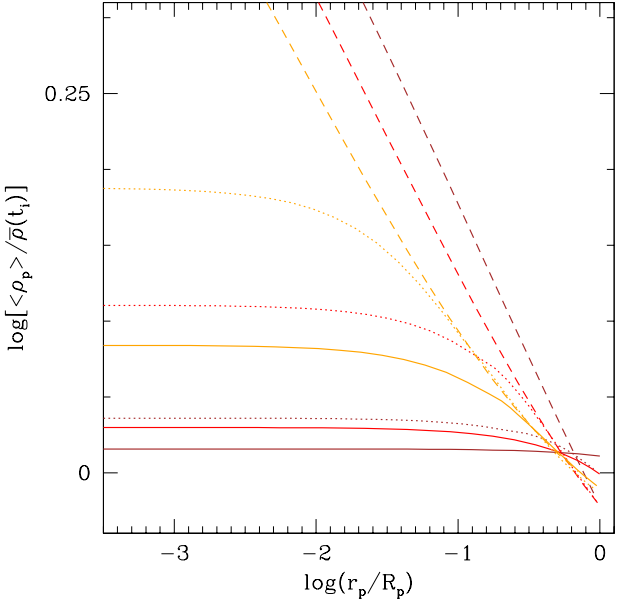}
\vskip -0.5 cm
 \caption{Spherically averaged (unconvolved) density profile for the
   same halo seeds and cosmologies as in Figure \ref{f1} (same
   lines).}
\label{f2}
\end{figure}

In Figure \ref{f2}, we show the spherically averaged density profile,
$\lav\rho\p\rav(r\p)$, of WDM (and CDM) halo seeds resulting from
equation (\ref{dp1}) for the peak trajectories depicted in Figure
\ref{f1}. As can be seen, contrarily to their CDM counterparts, the
WDM halo seeds show apparent flat cores with a universal mass very
approximately equal to the mass
$M\s=[4\pi/3]\,\bar\rho(t_0)\,(\lambda\s\eff/2)^{3}=4.7 \times 10^{5}$
\modot ($6.8\times10^{7}$ \modotc) for $m\X=10.8$ keV ($m\X=2$ keV)
sterile neutrinos associated with the effective free-streaming scale
length $\lambda\s\eff\equiv 2\pi/k\s\eff=\alpha$. Note that this is
different from the mass associated with scale length equal to $2\pi$
over the wavenumber where $T\s^2(k)$ decreases to 0.5, also often used
to estimate the free-streaming mass. This latter scale length is
substantially greater than $\alpha$, so it would correspond to the
scale length where the effects of WDM begin to be noticeable rather
than the minimum size of density perturbations. For this reason,
following \citet{SSMM11}, the mass $\sim 1.8\times10^{9}$ 
\modot ($\sim 1.9\times10^{11}$ \modotc) associated with it will be
called the half-mode mass, $M\hm$.

If the velocity dispersion in the seed were negligible, its flat core
with mass $\sim$$ M\s$ would expand and collapse at once, with no
shell-crossing before turnaround\footnote{In fact, these shells would
  not cross each other even after turnaround, meaning that such a flat
  perturbation would oscillate for ever and would not
  virialise. However, the subsequent shells coming from outside the
  flat core do cross them and the system finally
  virialises.}. Consequently, these flat protohalo cores with mass
$\sim$$M\s$ set a lower bound for the mass of WDM haloes. In contrast,
there is no lower bound for the mass of CDM haloes either. One can
find ancestors of any accreting halo with $M$ at $t$ down to
arbitrarily small cosmic times and with arbitrarily low masses.

\section{WDM Halo Density Profile}\label{warm}

The lower-bound mass, $M\s$, for haloes in the WDM cosmology just
mentioned arises from the spectrum cutoff. But this bound mass has
been obtained by neglecting the WDM particle velocity dispersion,
$\sigma\DM$. Actually, the non-negligible velocity dispersion of WDM
particles causes the total energy $E\p(R\p)$ in the seed
(eq.~[\ref{E1}]) to become positive for radii $R\p$ below some value
$R\E$. This means that shells inside that radius, with velocity
dispersion larger than the Hubble velocity, expand more rapidly than
outer ones, which causes the system to run out of linear regime,
leading to the formation of caustics. These caustics will fragment
(owing to the perturbation of the surrounding matter) and give rise to
small nodes with masses substantially less than $M\E\equiv M(R\E)$,
the mass of that part of the seed with null total energy. This means
that haloes with masses below $M\E > M\s$ do not actually evolve in
the bottom-up fashion and that the typical WDM halo density profile
will not satisfy equation (\ref{mr2}) at radii enclosing $M\E$.

\begin{figure}
 \includegraphics[scale=0.44]{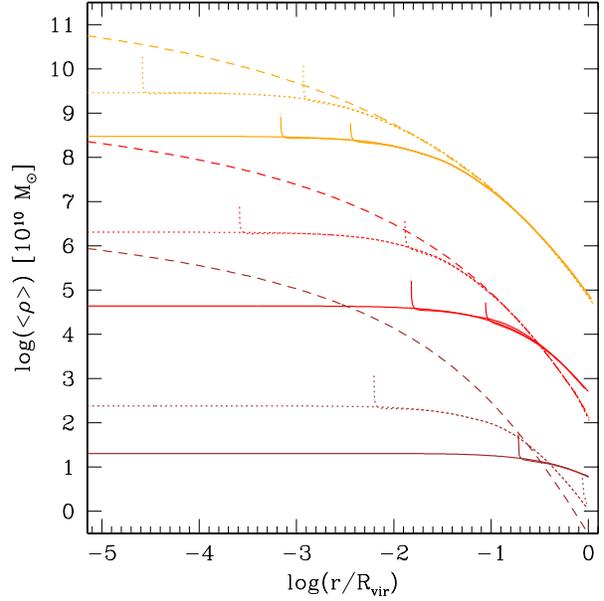}
\vskip -0.5 cm
 \caption{Typical spherically averaged density profiles predicted for
   the same haloes and in the same cosmologies (same lines) as in
   previous Figures, with null velocity dispersion (curves down to the
   halo centre), with thermal velocity dispersion (down to the inner
   break) and with non-thermal velocity dispersion (down to the outer
   break). To avoid crowding the profiles corresponding to $10^9$
   \modot and $10^{13}$ \modot have been shifted 2.5 dex downwards and
   upwards, respectively.}
\label{f3}
\end{figure}

According to \citet{Bo08}, the value of $\sigma\DM$ today
for non-thermal sterile neutrinos is related to the free-streaming
wavenumber through
\beq
k\s\eff\approx \left[\frac{3H^2(t_0)\,\Omega\M}{2\sigma\DM^2(t_0)}\right]^{1/2}\,.
\eeq
In the present case, this leads to $\sigma\DM(t_0)=1.09$ km s$^{-1}$
(0.23 km s$^{-1}$). This value is markedly greater than the one
usually adopted in this kind of studies, which might overestimate the
effects of the velocity dispersion on the predicted structure of WDM
haloes. For this reason, we will be more conservative and adopt
$\sigma\DM(t_0)=0.075$ km s$^{-1}$. To study the effects of changing
the value of $\sigma\DM(t_0)$, the density profiles so obtained will
be compared to those arising from null $\sigma\DM(t_0)$, so as to see
the effects of the cutoff in the spectrum alone, as well as from a
value of $\sigma\DM(t_0)$ equal to that of thermal neutrino-like WDM
particles. According to \citet{St06}, this latter velocity
dispersion is given by,
\beq 
\sigma\DM^3(t_0)=0.042^3\!\left(\!\frac{h}{0.7}\!\right)^{\!\!2}\!\left(\frac{m\X}{1~{\rm keV}}
\right)^{\!-4}\!\frac{\Omega\WDM}{0.273}\left({\rm km\;s^{-1}}\right)^3, 
\eeq
leading for $m\X=10.8$ keV ($m\X=2$ keV) particles to $0.015$ km
s$^{-1}$ ($\sigma\DM(t_0)=0.0018$ km s$^{-1}$). The reference
$\Lambda$CDM halo profiles are derived assuming a negligible
$\sigma\DM(t_0)$.

In Figure \ref{f3}, we plot the typical spherically averaged halo
density profiles for current haloes with the same masses as used in
the previous Figures, each of them for the three values of
$\sigma\DM(t_0)$ just mentioned. All the profiles deviate from the
corresponding CDM halo profiles in the same monotonous way leading to
a flat core with mass $M\s$. However, the only profile that can be
strictly traced down to the halo centre is for null velocity
dispersion. The remaining profiles (for non-vanishing
$\sigma\DM(t_0)$) show a sharp up-turn to infinity at a small enough
radius. This reflects the fact that, for $M$ approaching the value
$M\E$ where $E\p(M)$ vanishes (eq.~[\ref{E1}]), equation (\ref{mr2})
is no longer valid. For thermal velocity dispersion, the minimum
radius reached essentially coincides with the edge of the flat core,
while, for larger values of $\sigma\DM(t_0)$, it is significantly
larger. The mass $M\E$ encompassed by the minimum radius reached in
the case of non-thermal velocity dispersion shows a slight trend to
diminish with increasing halo mass. Specifically, for $m\X=10.8$ keV
($m\X=2$ keV) and $\sim 0.075$ km s$^{-1}$, it is about $10^{9}$
\modot ($10^{10}$ \modotc), $1.8\times 10^8$ \modot ($4.6\times10^{8}$
\modotc) and $8.1\times 10^7$ \modot ($1.2\times10^{8}$ \modotc) for
haloes with $10^9$ \modotc, $10^{11}$ \modot and $10^{13}$ \modotc,
respectively. Given the values of $R\vir$, respectively equal to 0.026
Mpc, 0.12 Mpc and 0.57 Mpc, this leads to core radii of about 0.2 kpc
(2 kpc), 0.1 kpc (1 kpc) and 0.05 (0.5 kpc), respectively.

\section{Conclusions}\label{dis}

The typical spherically averaged density profile for haloes with
masses greater than $M\s$ in the WDM cosmologies can be derived
analytically by means of the SVMS model. The density profiles so
obtained show a clear flattening relative to the CDM profiles,
independent of the particle velocity dispersion, that evolves into a
flat core with mass equal to $\sim 4.7\times 10^7$ \modot ($4.1 \times
10^{5}$ \modotc) for $m\X=10.8$ keV ($m\X=2$ keV) sterile
neutrinos. This minimum mass agrees with the mass functions derived
from WDM models and simulations (e.g. \citealt{Zavea09,SM11}). For
$m\X \ge 10.8$ keV, this leads to core radii that are however less
than $\sim 1$ kpc as observed in LSB galaxies \citep{KK11,Saluea12},
in agreement with \citet{Macea12b}. The right core radii require
$m\X\la 2$ keV.

The only effect of particle velocities is that they prevent from
reaching the flat core. They cause protohaloes to have null total
energy within some small radius encompassing the mass $M\s$. This
produces caustics as protohaloes expand, leading to their
fragmentation into small nodes, which would explain the presence of
haloes with masses below $M\s$ in $N$-body simulations of WDM
cosmologies. Consequently, small mass haloes do not develop in the
bottom-up fashion and their density profile cannot be recovered by
means of the SVMS model. Specifically, for $\sigma\DM(t_0)=
  0.075$ km s$^{-1}$, the minimum radius that can be reached for
  $m\X=10.8$ kev ($m\X=2$ keV) sterile neutrinos is 26 kpc ($26$ kpc),
  $1.5$ kpc (12 kpc) and 0.7 kpc (2.3 kpc) in haloes with $10^9$
  \modotc, $10^{11}$ \modot and $10^{13}$ \modotc, respectively. In
  principle, the density profile inside that radius may not be flat as
  essentially found with null velocity. However, given the fixed
  radius and inner mass, it should not be very different either. The
  results of numerical simulations by \citet{Macea12b}
  confirm such expectations. The ``unresolved region'' defines the
  minimum mass of haloes formed hierarchically in the case of non-null
  velocity dispersion. It coincides with the mass of the whole halo
  for objects with $10^{9}$ \modot ($10^{10}$ \modotc) in the case of
  $m\X=10.8$ keV ($m\X=2$ keV) and $\sigma\DM(t_0)\sim 0.075$ km
  s$^{-1}$; for thermal velocities it is more than one order of
  magnitude less. Thus, the minimum mass of haloes grown
  hierarchically in the case of $m\X=2$ keV sterile neutrinos
  leading to cores with the right size is smaller than $10^9$
  \modotc. These results are slightly less restrictive than
  those found by \citet{Macea12b}.

%\acknowledgments
\vspace{0.75cm} \par\noindent
{\bf ACKNOWLEDGEMENTS}
\vspace{0.25cm} \par

\noindent This work was supported by the Spanish DGES,
AYA2009-12792-C03-01, and the Catalan DIUE, 2009SGR00217. One of us,
JV, was beneficiary of a Spanish FPI grant.

\end{document}